\documentclass[journal]{IEEEtran}
\usepackage[utf8]{inputenc}
\usepackage[T1]{fontenc}
\usepackage{lmodern}
\usepackage{amsmath,amssymb,amsfonts}
\usepackage{graphicx}
\usepackage{subcaption}
\usepackage{float}
\usepackage{url}
\usepackage{cite}
\usepackage{geometry}
\usepackage{csquotes}
\title{\textbf{Prompt-to-Primal Teaching}}

\author{
Euzeli C. dos Santos Jr.\\
\textit{Purdue University, Indianapolis, IN, USA}\\
\texttt{edossant@purdue.edu}
}

\date{Manuscript submitted October 2025}

\begin{document}
\maketitle

\begin{abstract}
This paper introduces Prompt-to-Primal (P2P) Teaching, an AI-integrated instructional approach that links prompt-driven exploration with first-principles reasoning, guided and moderated by the instructor within the classroom setting. In P2P teaching, student-generated AI prompts serve as entry points for inquiry and initial discussions in class, while the instructor guides learners to validate, challenge, and reconstruct AI responses through fundamental physical and mathematical laws. The approach encourages self-reflective development, critical evaluation of AI outputs, and conceptual foundational knowledge of the core engineering principles. A large language model (LLM) can be a highly effective tool for those who already possess foundational knowledge of a subject; however, it may also mislead students who lack sufficient background in the subject matter. Results from two student cohorts across different semesters suggest the pedagogical effectiveness of the P2P teaching framework in enhancing both AI literacy and engineering reasoning.
\end{abstract}

\begin{IEEEkeywords}
Engineering education, learning with AI, and teaching with LLM.
\end{IEEEkeywords}

\section{Introduction}
Artificial intelligence (AI) and large language models (LLMs) such as ChatGPT and Gemini have become widespread tools in education, offering instant access to complex explanations, examples, and derivations. However, while such tools can support learners with existing conceptual frameworks, they often mislead novices who lack the domain knowledge to evaluate AI outputs critically. Technical literarure establishes empirical support for integrating AI into pedagogical tools \cite{lee2025,carrasco2025,qian2025}, which includes adaptive paths, personalized testing, and real-time analysis \cite{ward2025}. As a result of this integration the authors noticed gains in GPA and reduced study hours experienced by the students. Another research finds that Generative AI offers a transformative approach to Engineering Education by facilitating a more interactive, personalized, and adaptive curriculum \cite{ciolacu2024}. On the other hand, \cite{guedes2025} concludes that the use of AI must be balanced by a proactive effort to address serious risks like algorithmic bias, privacy, and job displacement. 

In engineering education, where problem-solving relies on rigorous derivations, physical constraints, and mathematical reasoning, uncritical dependence on AI risks producing “surface-level understanding” \cite{hao2025}. Existing teaching frameworks such as Just-in-Time Teaching (JiTT) \cite{novak1999} and Flipped Classroom models \cite{bishop2013} emphasize pre-class preparation and active engagement but rarely address how to integrate AI responsibly in ways that preserve rigorous reasoning.

To fill this gap, this paper proposes a new framework called Prompt-to-Principal Teaching (P2P), designed to align AI-assisted exploration with first-principle validation. In this approach, learning begins with student-generated prompts, queries posed to AI models, followed by instructor-led reasoning grounded in first principles. The aim is not to replace traditional instruction with AI assistance but to use AI as a pedagogical catalyst for inquiry, reflection, and conceptual grounding.

\begin{figure}[b]
\centering
\includegraphics[width=1.0\linewidth]{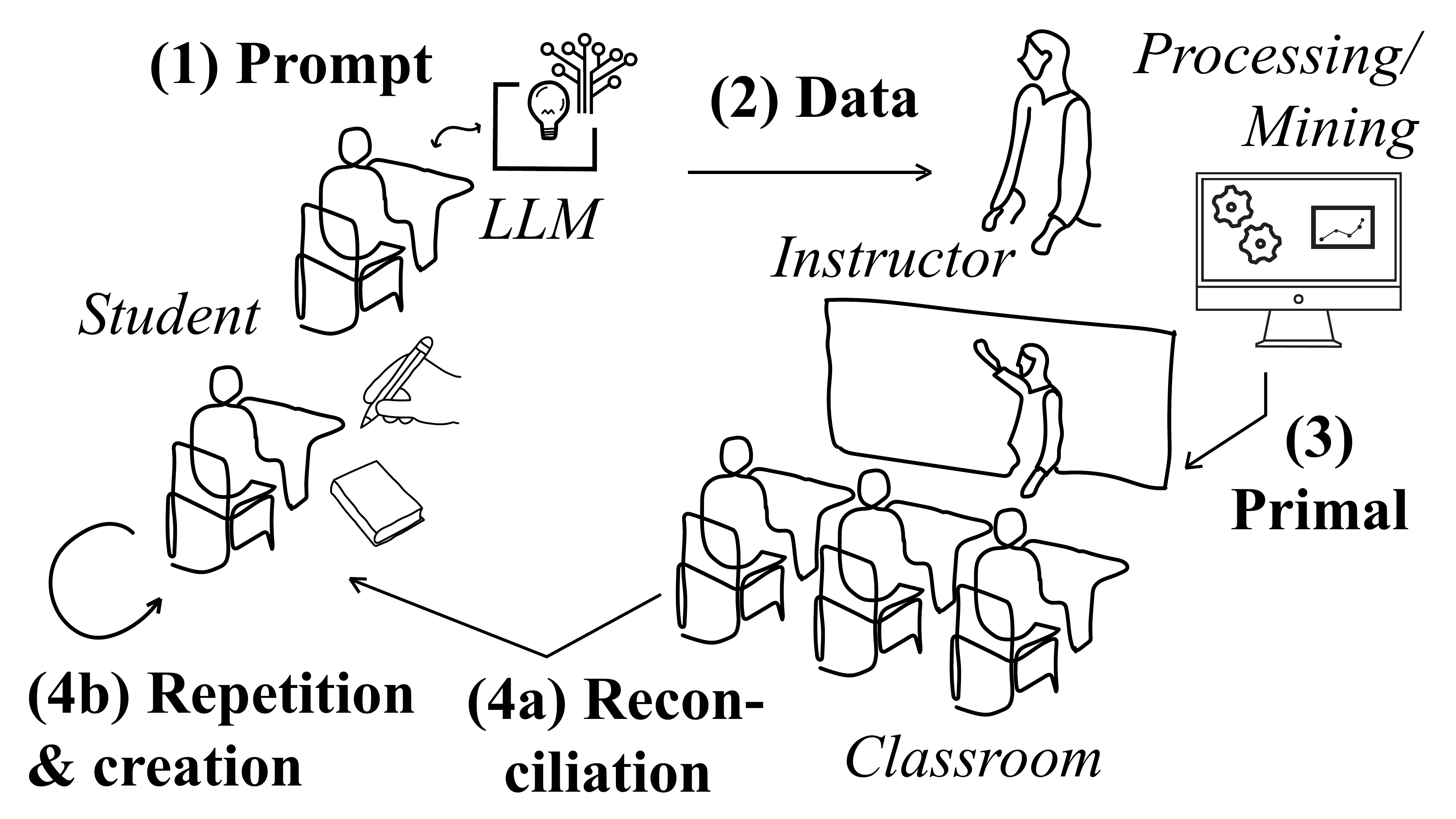}
\caption{Schematic representation of the Prompt-to-Primal (P2P) teaching framework, illustrating the cyclical integration of AI-based exploration, instructor-mediated analysis, first-principles grounding, and reflective repetition leading to creative application.}
\label{fig:framework}
\end{figure}

Fig. \ref{fig:framework} illustrates the conceptual structure of the P2P teaching framework, which integrates AI–assisted exploration with first-principles reasoning in a cyclic pedagogical model. The process begins with the Prompt phase (1), where the student interacts with a Large Language Model (LLM) to generate exploratory dialogues around a given topic. The Data phase (2) follows, as the instructor analyzes these dialogues to extract misconceptions, knowledge gaps, and thematic patterns that inform classroom instruction (AI is also used by the instructor to process and mine data). In the Primal phase (3), the instructor leads students in reconstructing or validating the AI-generated explanations through grounding in first principles, ensuring conceptual rigor.

Finally, Phase 4, which encompasses Reconciliation (4a) and Repetition/Creation (4b), closes the learning loop: students critically reflect on discrepancies between AI-derived and principle-based knowledge, then revisit and reapply the concepts to reinforce understanding by repeating first principles presented in class, which is in turn done by solving problems and/or implementing either simulation, experimentation or solving related problems. Fig. \ref{fig:framework} encapsulates the iterative and reflexive nature of the P2P model, where AI-driven inquiry, foundational reasoning, and hand-writting repetition converge to promote deeper, more durable learning. 

\begin{figure*}[h]
\centering
\begin{subfigure}[b]{0.4\linewidth}
    \centering
    \includegraphics[width=\linewidth]{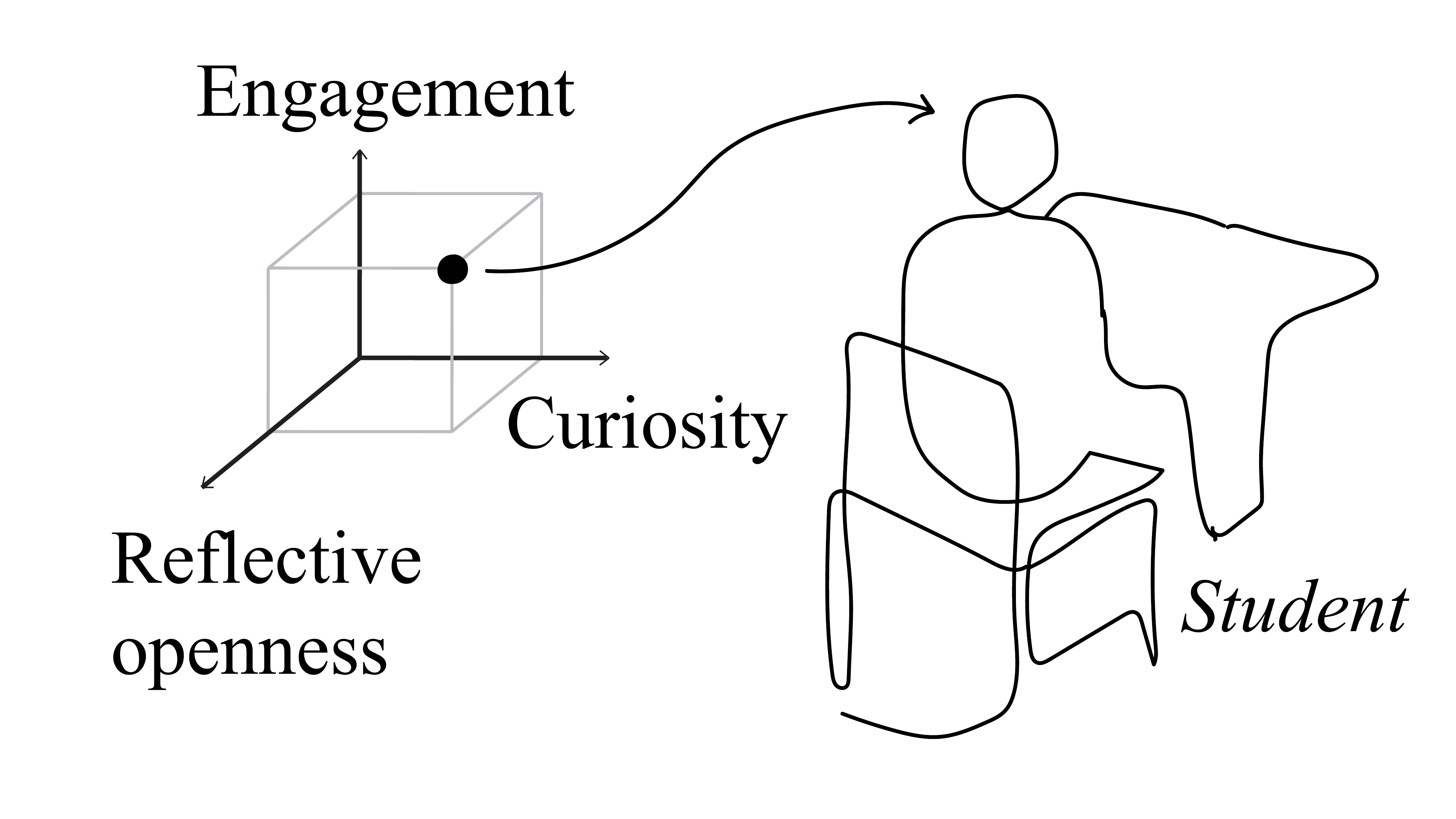}
    \caption{}
\end{subfigure}
\begin{subfigure}[b]{0.5\linewidth}
    \centering
    \includegraphics[width=\linewidth]{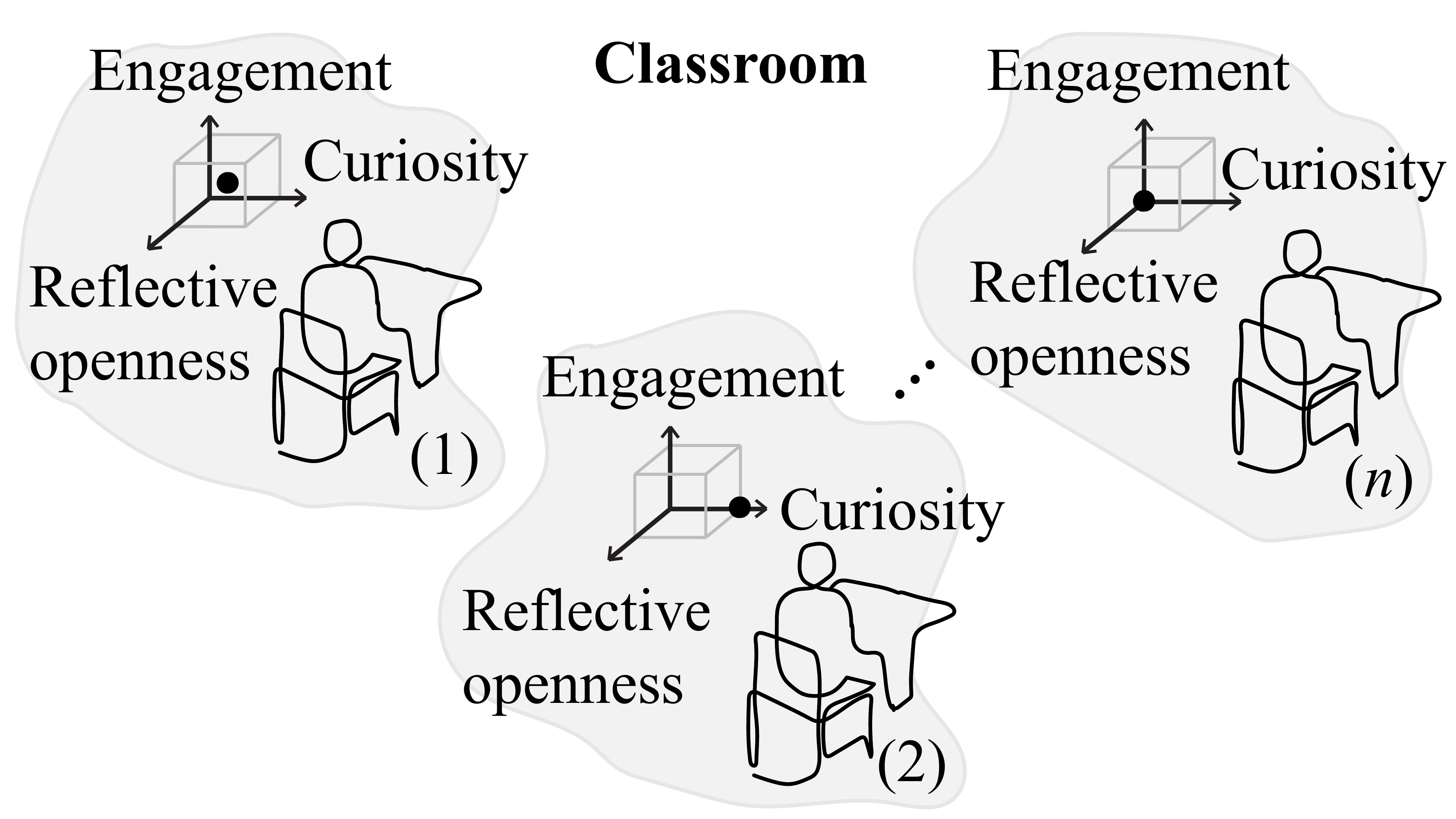}
    \caption{}
\end{subfigure}
\caption{Core dispositions (dimensions) contributing to learner efficacy in the formal educational context: (a) ideal situation when the student exhibit satisfactory levels of curiosity, engagement, and reflective openness; (b) a more realistic classroom with a diverse disposition.}
\label{fig:dispositions}
\end{figure*}

Fig. \ref{fig:dispositions} provides a conceptual illustration designed to represent the student’s perceived capability within the broader teaching–learning framework. The diagram serves as a visual element for understanding how students’ internal dispositions interact with the learning environment. Specifically, it delineates and grounds the three fundamental dispositional dimensions that shape the learner’s experience: (1) Curiosity, which drives the intrinsic motivation to explore new ideas; (2) Engagement, reflecting the sustained cognitive and emotional investment in learning activities; and (3) Reflective Openness, denoting the willingness to reconsider assumptions and integrate new perspectives. Together, these elements constitute a holistic model of how students perceive and enact their agency in educational contexts. These dispositions are schematically rendered as a three-dimensional cognitive-behavioral manifold. Fig. \ref{fig:dispositions}(a) illustrates the ideal dispositional state, wherein the learner's profile is situated at a high coordinate across all three axes, representing the comprehensive integration of motivational, active, and metacognitive \cite{karagianni2024} elements needed for potentially achieving intellectual growth and sustaining the identified virtuous cycle of learning. Conversely, Fig. \ref{fig:dispositions}(b) depicts a realistic pedagogical scenario by mapping the diverse dispositional profiles inherent to a classroom setting. This clearly illustrates the common heterogeneity of these core attributes across the cohort. As discussed throughout this paper, Phase 1 of the P2P teaching framework (see Fig. 1) plays a central role in fostering the dimension of curiosity, serving as the initial stimulus for exploratory learning through student–AI interaction. Phases 2 and 3 are primarily oriented toward promoting engagement, as they enable the instructor to design and conduct classroom activities informed by the data collected from student–LLM dialogues, thereby addressing the specific learning needs of that cohort. This concept is borrowed from the Just-in-Time Teaching method \cite{novak1999}. The Reconciliation phase (Phase 4a) is essential for cultivating reflective openness, guiding students to critically compare AI-generated insights with principle-based reasoning. Finally, the Repetition and Creation Phase 4b consolidates and extends learning by reinforcing conceptual understanding and stimulating creative application, ensuring durable and meaningful knowledge acquisition across all students.

\section{The P2P Teaching Framework}

\subsection{Phase 1 – Prompt (Exploration through AI)}

This approach integrates a pre-class activity where students initiate the learning process by developing AI prompts derived from the scheduled lecture content. This instructional design parallels the methodology of Just-in-Time Teaching (JiTT) \cite{novak1999}. However, within the proposed P2P teaching model, the students are responsible for prompt generation. Rather than furnishing pre-determined questions, the instructor provides an overarching theme or disciplinary topic to foster an exploratory dialog between the student and the AI tool. For example, the instructor can introduce the foundational concept, such as: “Explain the operation of a controlled voltage source for DC machines,” and follow it with a more focused constraint, such as: \enquote{I wanted to know about regulating voltage with a control signal.}

Subsequent inquiries are student-driven and not constrained by pre-established prompts; indeed, students are explicitly encouraged to engage in an unrestricted, iterative conversation with the Large Language Model (LLM) concerning the subject matter under consideration. Students are required to save the dialog and subsequently present it to the instructor a day prior to the beginning of the class. Whenever feasible, the “conversation” between the student and the LLM should be conducted through spoken interaction rather than typing, as verbal communication fosters a more natural exchange and allows the learner’s curiosity to emerge more naturally.

\begin{figure}[b]
\centering
\includegraphics[width=1.0\linewidth]{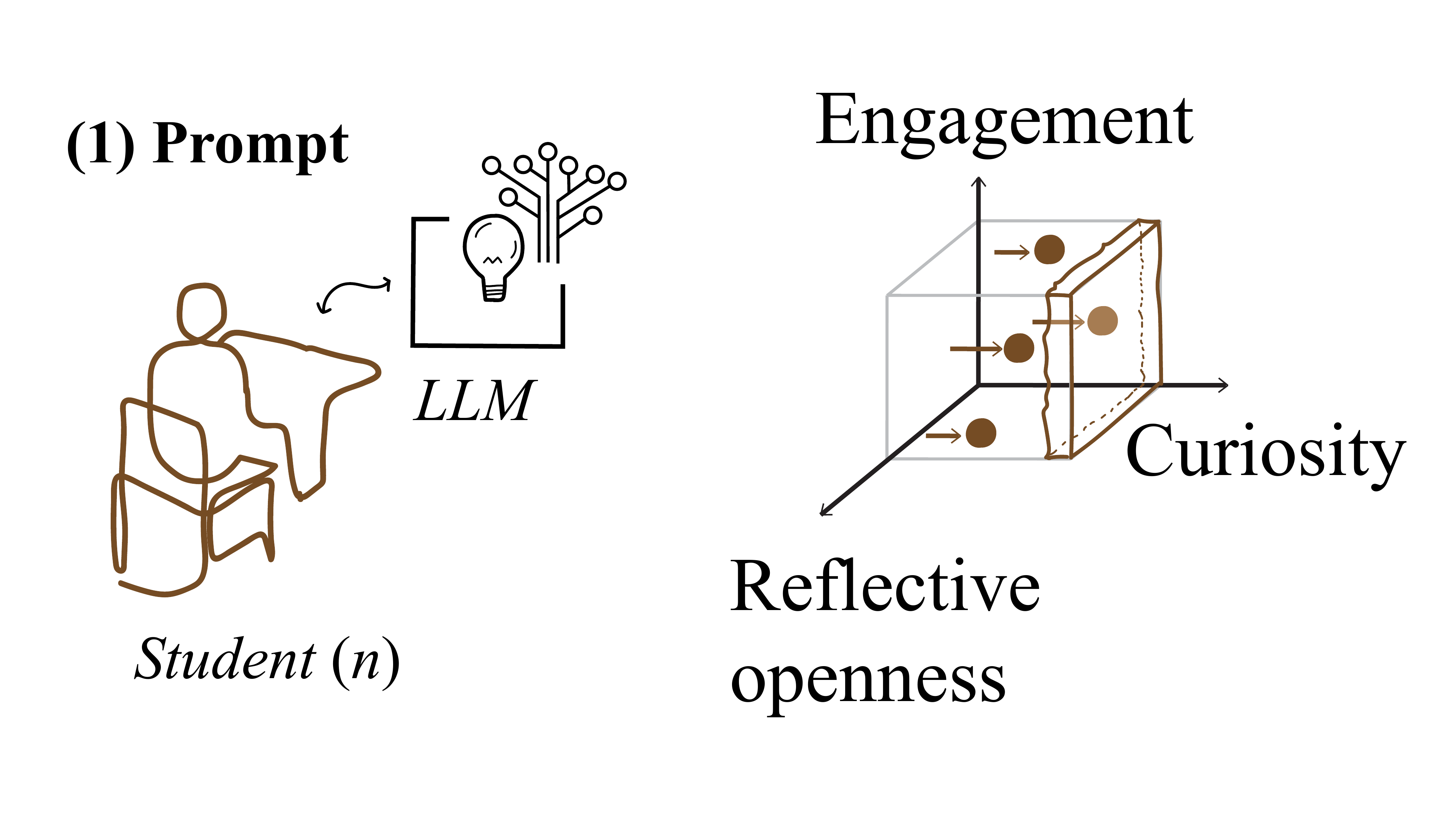}
\caption{Schematic illustrating the pre-class activity where a student initiates an iterative dialogue with a Large Language Model (LLM) via self-generated prompts, which serves to foster the core learning disposition of Curiosity and generate exploratory data for the instructor.}
\label{fig:phase1}
\end{figure}

Fig. \ref{fig:phase1} illustrates how the Prompt aims to foster the core disposition of Curiosity in the cohort. The dots and vectors pointing toward a higher coordinate on the Curiosity axis, suggests that the students interaction serve as the initial stimulus for inquiry-based learning, leading to dialogues that the instructor will later analyze to inform classroom instruction. The dots in this figure represent a diverse range of students with coordinates spanning the low to high spectrum for both Engagement and Reflective openness.

\subsection{Phase 2 - Data (Processing and Mining Using AI)}

The instructor will process the data and find statistic meaning from the conversation between students and LLM. This involves a rapid, pre-class analysis of the collected dialogues to identify several critical pedagogical elements. Specifically, the instructor will look for common misconceptions introduced by the AI or reinforced by student inquiry, emergent points of conceptual confusion that require immediate classroom attention, and the range of sophistication in student-generated prompts. This rapid diagnostic process transforms the raw dialogue data into actionable anchors for the next in-class session. By synthesizing the students’ interactions with the LLM, the instructor gains a real-time, high-fidelity map of the class's prior knowledge and gaps, allowing them to tailor the subsequent instruction to directly address any possible \enquote{illusion of understanding} and ensure that the core principles are introduced precisely where the AI has provided plausible but physically flawed explanations. The resulting insights dictate which fundamental first principles will be used to validate, challenge, or correct the AI's output in the next phase, thus moving the lesson from prompt-driven exploration to first-principles validation.

\subsection{Phase 3 – Primal (Grounding through First Principles)}

This phase represents the core of the P2P instructional methodology: the exploration initiated by the LLM is the starting point that directs how all subsequent phases will be executed. Using the diagnostic insights gathered in Phase 2, the instructor strategically leads the class in a collective critical examination of the AI-generated responses. After discussing the most common themes from the collected data, the next activity involves selecting key claims (even in the absence of statistical significance) or derivations from student dialog, particularly those containing subtle errors or superficial explanations, and subjecting them to formal domain-specific validation.

This active reconstruction serves a dual purpose: it directly addresses the identified conceptual vulnerabilities by demonstrating how established principles (the enquote{Primal} knowledge) supersede the LLM's pattern-matching, and it fosters the development of adaptive expertise. By observing the instructor model this critical process (how to triangulate an answer against fundamental truths) students internalize the method of inquiry, transforming the AI from an uncritical source of answers into a pedagogical catalyst for deeper, more resilient engineering reasoning. 

\begin{figure}[h]
\centering
\includegraphics[width=1.0\linewidth]{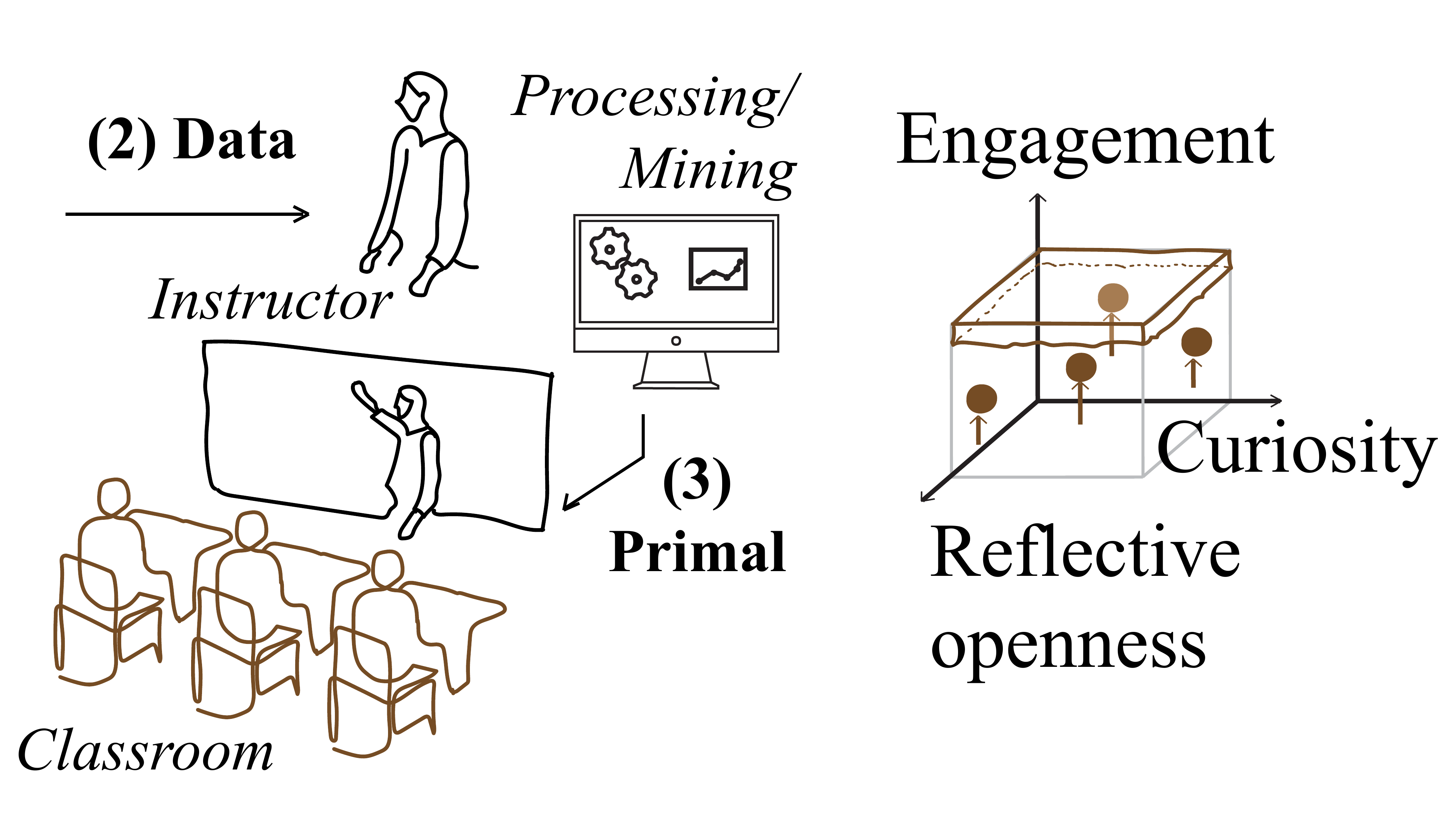}
\caption{Phases 2 and 3 - Data and Primal (Promoting Engagement). Schematic demonstrating the synergistic role of Phase 2 (Instructor Data Processing and Mining) and Phase 3 (Classroom Grounding via First Principles) in the P2P framework.}
\label{fig:phase23}
\end{figure}

Fig. \ref{fig:phase23} illustrates Phase 2: Data (Processing and Mining) and Phase 3: Primal (Grounding through First Principles) of the Prompt-to-Primal (P2P) teaching framework. This figure emphasizes how these phases work together to promote the core disposition of Engagement in the cohort. The final two components of the P2P framework, sub-elements 4a (Reconciliation) and 4b (Repetition and Creation), are presented next and are fundamentally dependent on the student's active engagement to effectively close the learning cycle.

\subsection{Phase 4a – Reconciliation (Critical Reflection)}

This phase is designed to solidify learning by establishing an explicit comparison between the AI's, pattern-based output and the instructor-led, first-principles derivation. This crucial step moves students beyond merely identifying the type of information received to analyzing the source and nature of the information, thus completing the pedagogical loop from exploration (Prompt) to grounding (Primal). It directly combats the \enquote{illusion of understanding} by forcing students to articulate the differences between fluency and expertise. 

The Reconciliation phase is designed to achieve several important pedagogical goals, primarily by promoting reflection on their learning process, contrasting their initial, AI-mediated understanding with the rigorously grounded understanding derived from first principles, thereby learning how knowledge is validated. Simultaneously, it fosters AI literacy by cultivating a healthy skepticism, teaching students that while the LLM is an excellent tool for generating initial ideas, human-center rigorous analysis remain essential for validation and accuracy. Furthermore, this phase reinforces foundational knowledge; the act of comparing and contrasting strengthens the relevant neural pathways \cite{marzano2001}. As shown in Fig. \ref{fig:framework}, Phase 4a and Phase 4b require student commitment and prohibit the use of AI, especially when proceeding to the Repeating phase, as detailed next. 

\subsection{Phase 4b – Repetition and Creation (Constructive Application)}

Repetition serves as a vital consolidating mechanism within the P2P teaching framework, ensuring that the knowledge reconstructed through first-principles reasoning becomes deeply internalized and readily transferable. After students have critically examined, validated, and reconciled the information obtained from both LLM and first principles, revisiting the same concepts allows them to reinforce neural pathways associated with accurate understanding while extinguishing misconceptions introduced during earlier exploratory phases. It is important to note that Phases 1, 2, and 3 demand minimal time and effort from students, whereas the substantive learning effort and cognitive consolidation occur primarily during Phases 4a and 4b, especially with deliberate repetition, which transforms temporary comprehension into durable knowledge. Moreover, by re-engaging with the material through iterative discussion, problem-solving, or re-explanation, students strengthen their cognitive structures and develop automaticity in applying foundational concepts. 

The Creation phase concludes the P2P teaching cycle by guiding students from analytical reflection to productive synthesis and innovation. Having validated AI-generated content through first-principles reasoning, learners now apply their conceptual and methodological insights to higher-order tasks such as problem-solving, design, simulation, and experimentation.  

\begin{figure}[b]
\centering
\includegraphics[width=1.0\linewidth]{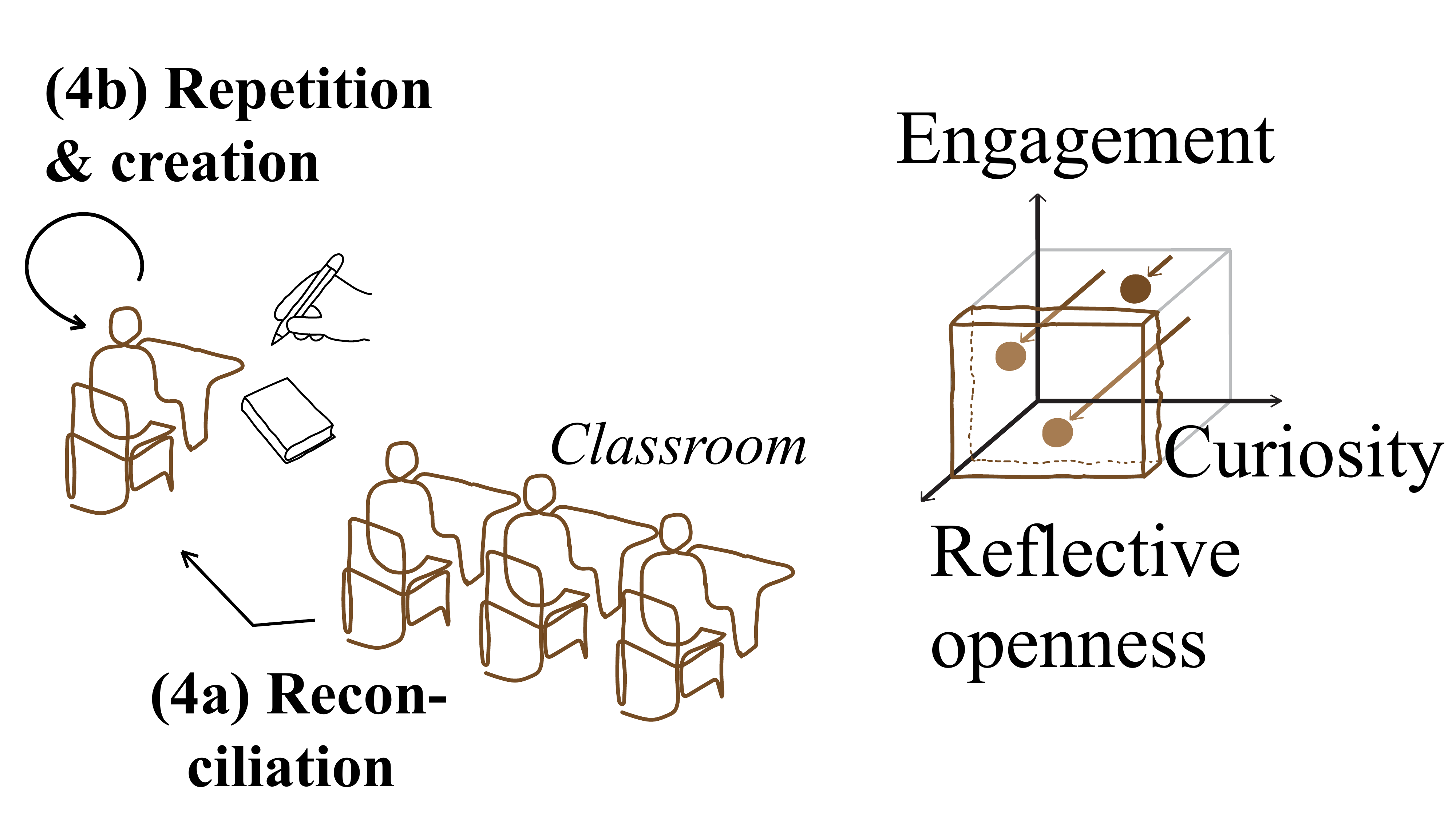}
\caption{Phase 4 - Reconciliation, Repetition and Creation (Cultivating Reflective Openness). Schematic illustrating the final, student-driven phases of the P2P framework, where the learner critically reconciles AI-generated content with first-principles knowledge (4a), then consolidates and applies that knowledge to new creative tasks (4b).}
\label{fig:phase4}
\end{figure}

Fig. \ref{fig:phase4} illustrates Phase 4: Reconciliation (4a) and Repetition - Creation (4b), which together represent the final, student-driven closure of the Prompt-to-Primal (P2P) teaching cycle. This phase is primarily designed to foster the core learning disposition of Reflective openness, but it depends on the student's commitment to carrying out these last two phases of the process.

\section{Rationale and Theoretical Foundation}

Learning in the age of AI presents both opportunities and challenges that demand a careful pedagogical approach. Four foundational pillars are presented in this section as the cornerstones of the proposed approach: \emph{(a) The Illusion of Understanding}, \emph{(b) Constructivist and Reflective Basis}, \emph{(c) Anchoring Knowledge in First Principles}, and \emph{(d) Repetition Works!}.

In today’s educational landscape, superficial fluency often conceals a lack of deep understanding, particularly when AI is employed in isolation. However, structured and theory-informed methodologies can foster enduring and meaningful learning. While AI tools can inadvertently create the illusion of understanding by delivering immediate, coherent solutions, constructivist frameworks emphasize active engagement, reflection, and learner-centered knowledge construction to counteract this passivity. Anchoring knowledge in first principles provides stable and foundational reference points that connect abstract theories to real-world applications, preventing the formation of detached from reality knowledge. Finally, strategically structured repetition, including spaced retrieval, interleaving, and the introduction of desirable difficulties, reinforces understanding and promotes transfer to novel problems. The synthesis of these interrelated concepts into a unified teaching approach is only feasible in the contemporary context, given the widespread accessibility of Artificial Intelligence.

\subsection{The Illusion of Understanding}

Learning a subject with the aid of AI tools can create a subtle, yet significant, \enquote{illusion of understanding}. When students rely heavily on AI to generate summaries, solve complex problems step-by-step, or explain concepts, the immediate availability of correct and coherent information can mask a fundamental lack of deep, internalized knowledge \cite{zhai2024}. The learner is passively consuming an answer rather than actively engaging in the cognitive struggle necessary for true comprehension, such as synthesizing information, troubleshooting errors, and forming independent critical judgments \cite{chi2009}. This reliance allows them to mistake fluency with AI-generated explanations for genuine mastery, leading to brittle knowledge that fails when the AI crutch is removed in unassisted application or assessment. This superficial learning bypasses the crucial process of building robust, durable understanding through effortful retrieval and generative practice \cite{messeri2024}.

AI systems operate through pattern recognition rather than understanding. They produce statistically coherent outputs but lack causal reasoning. Thus, expert users who possess strong background can triangulate the AI’s suggestions, which can be otherwise difficult for someone who just got exposed to that subject. This asymmetry motivates a structured, instructor-mediated framework, which is proposed in this paper with the P2P teaching.

\subsection{Constructivist and Reflective Basis}

In educational practice, constructivist theory transforms the role of the teacher from that of a knowledge transmitter to a facilitator or guide who designs environments that stimulate exploration and inquiry \cite{brooks1999}. Learning activities under this model often emphasize problem-solving, experimentation, and dialogue, allowing students to construct personal interpretations rather than memorize fixed answers. Tools such as project-based learning, peer instruction, and guided discovery are natural extensions of constructivist thinking. Moreover, constructivism supports differentiated learning pathways, recognizing that student's understanding happens through diverse cognitive frameworks shaped by their backgrounds and experiences. The theory has profound implications for modern pedagogical design, particularly in STEM and engineering education, where conceptual understanding and adaptive reasoning are crucial. By engaging learners in active sense-making, constructivist teaching not only deepens comprehension but also cultivates the critical and metacognitive skills necessary for lifelong learning in an ever-evolving knowledge landscape \cite{prince2006}.

Indeed, the emphasizes on learner-centered knowledge construction through active engagement and reflection, aligns closely with the personalized and interactive capabilities afforded by AI. AI-driven educational systems can dynamically adapt to individual learners’ cognitive states, prior knowledge, and learning trajectories, thereby facilitating the scaffolding processes that are central to a constructivist pedagogy. Through adaptive feedback, conversational interfaces, and context-aware guidance, AI can enable students to engage in iterative sense-making rather than passive content absorption. This synergy transforms learning into a dialogic process, where the learner co-constructs understanding in response to real-time, data-informed interactions, thus operationalizing constructivist principles in a technologically enhanced educational environment.

The P2P teaching approach draws upon constructivist learning theory, where students construct knowledge through active interaction and reflection, and awareness-based learning. The inclusion of AI as a dialogic partner fosters inquiry-based learning while the instructor ensures coherence between student understanding and first principles, as presented next.

\subsection{Anchoring Knowledge in First Principles}

Anchoring Knowledge in First Principles is a pedagogical approach in engineering education that explicitly links abstract, fundamental theories (first principles) to concrete, real-world engineering problems or case studies. First principles are the core, self-evident propositions and foundational laws of physics, mathematics, and science that engineering is built upon, such as Newton's laws or the principle of conservation of energy. The anchoring methodology, rooted in situated cognition and experiential learning theory, addresses the common challenge of inert knowledge, where students can recall theoretical concepts but struggle to apply them to novel, complex situations encountered in professional practice \cite{wilson2000}. By deeply contextualizing these fundamental principles, educators provide a stable reference point that students can continuously return to, ensuring the theoretical knowledge is perceived as a \enquote{tool} for more complex problems rather than just content to be memorized for an exam.

\subsection{Repetition Works!}

Modern research shows that repetition enhances learning only when it is structured strategically rather than performed mindlessly \cite{saville2011}. Techniques such as spaced practice, where study sessions are distributed over time, and retrieval practice \cite{agarwal2021}, where learners actively recall information, produce stronger and longer-lasting memory than simple massed repetition. Other strategies, such as interleaving different topics or problem types and introducing desirable difficulties, make repetition more effective by encouraging effortful processing, strengthening memory consolidation, and improving the ability to transfer knowledge to new contexts. 

Repetition also serves as a mechanism for stabilizing cognitive representations, allowing new information to transition from short-term awareness to long-term mastery. In this sense, repetition functions not as mechanical drill but as a scaffold for conceptual integration, bridging the gap between initial comprehension and adaptive, expert-level performance. Also, the literature indicates that handwriting uniquely enhances learning and memory by promoting widespread connectivity across key brain regions, including the parietal and central areas, whereas typing on a keyboard does not produce the same neural effects \cite{vanderweel2024}. 

\section{Application Example - Electromechanical Motion Devices}

To illustrate the implementation of the Prompt-to-Primal (P2P) teaching framework, this section presents an application focused on a class of the course Electromechanical Motion Devices, specifically the topic of Permanent Magnet DC (PMDC) machine and its control mechanisms. The instructional process followed the four P2P phases.

\subsection*{Phase 1 – Prompt (Exploration through AI)}
Before the lecture, students were instructed to engage in a conversation with an LLM about the topic “Permanent Magnetic DC Machine – Control Strategies.” They were advised to ask any question that emerged naturally, without concern for correctness, and to document the full dialogue. This activity aimed to stimulate curiosity and autonomous inquiry, encouraging students to explore topics such as application, modeling, or any other aspect related to this subject. As expected, AI provided technically plausible but occasionally inconsistent with the background of the students.

\subsection*{Phase 2 – Data (Processing and Mining)}

After collecting transcripts of the student–LLM dialog, the instructor performed a rapid qualitative analysis. The findings guided the instructional design of the following class, allowing the instructor to tailor the lecture around these conceptual gaps. 

The following data was processed through AI-assisted text-mining:
\begin{enumerate}
    \item total questions asked: 29,
    \item unique question intents: 21 (72.4\%),
    \item repeated questions (same intent asked): 8 (27.6\%),
    \item repeated question groups that produced the same answers (same facts, different depth/wording): 100\%, and
    \item duplicate-intent sets with conflicting answers: 0\%.
\end{enumerate}

The students cared about the following topics:
\begin{enumerate}
    \item PMDC vs wound-field DC (differences/field-weakening) 10.3\%, PWM \& power conversion (what is PWM, choppers/H-bridge, braking modes) 27.6\%,
    \item Modeling \& equations (electrical/mechanical, back-EMF, torque) 20.7\%,
    \item Control architecture (loops/sensors/closed-loop basics) 17.2\%,
    \item implementation details (PWM, sample-rate sync, tuning, Simulink) 13.8\%,
    \item assignments/examples/meta (circuit example, homework, ``what to ask'') 6.9\%, and
    \item practical issues/limitations (saturation, noise, thermal, wear) 3.4\%.
\end{enumerate}

\subsection*{Phase 3 – Primal (Grounding through First Principles)}
The processed data collected from Phase 2 will be used to determine the directions on how to present the first principles. Based on the statistical analysis, the lecture should strategically emphasize several key areas to address the main sources of students inquiries and interest. A primary focus should be placed on differentiating permanent-magnet direct current (PMDC) machines from a DC motor with field circuitry. In addition, the lecture should ensure that students gain confidence in using the two fundamental relationships governing PMDC behavior: torque being proportional to armature current and induced voltage being proportional to angular speed. These linear equations form the basis of the motor’s controllability. Another major area deserving instructional time is the power electronics interface, i.e., Pulse Width Modulation (PWM) and the operation of the H-bridge converter. Since amost 28\% of the questions centered on these topics, students would benefit from a clear explanation of those topics. The concept of cascaded control accounted for close to 18\% of students' questions. However, it falls outside the scope the 300-level class and therefore was not presented. 

\subsection*{Phase 4a – Reconciliation (Critical Reflection) / Phase 4b – Repetition \& Creation (Constructive
Application)}
Subsequently, students completed a brief survey to compare their AI-derived understanding against the validated model developed from first principles in the classroom. The survey asked the following questions:

1. Conceptual Alignment

\noindent
After the lecture, how closely did your AI-based understanding of Permanent Magnet DC Machines align with the model derived from first principles in class?

\noindent
(\ ) Completely aligned \ \ (\ ) Mostly aligned \ \ (\ ) Partially aligned \ \ (\ ) Mostly different \ \ (\ ) Completely different

2. Depth of Understanding

\noindent
How much did reconciling the AI explanation with the first-principles model help you refine your conceptual understanding?

\noindent
(\ ) Significantly improved \ \ (\ ) Moderately improved \ \ (\ ) Slightly improved \ \ (\ ) No change

3. Critical Validation Skill

\noindent
After this experience, how confident do you feel in evaluating whether an AI-generated technical explanation respects fundamental physical principles?

\noindent
(\ ) Very confident \ \ (\ ) Somewhat confident \ \ (\ ) Neutral \ \ (\ ) Slightly uncertain \ \ (\ ) Not confident

4. AI as Learning Partner

\noindent
To what extent do you now view AI as (a) a useful exploratory tool, or (b) a source that requires systematic verification through first-principles reasoning?

\noindent
(\ ) Mostly exploratory \ \ (\ ) Balanced view \ \ (\ ) Mostly requires verification \ \ (\ ) Not useful without verification

4. Subsequent phase

\noindent
Do you agree to independently revisit the class's first principle by writing it out manually, excluding the use of AI?

\noindent
(\ ) yes \ \ (\ ) no \ \ (\ ) maybe

The Repetition Phase presents a challenge in assessment, as it relies primarily on the students’ individual commitment to revisiting and internalizing the first-principles discussed in class. This stage encourages learners to reproduce key derivations and reasoning processes by hand, reinforcing conceptual depth through deliberate practice. In contrast, the Creation Phase, which emphasizes problem-solving and the application of learned principles, can be quantitatively evaluated through student performance on homework assignments and examinations, providing measurable evidence of knowledge integration and transfer.

\section{Assessment and Learning Outcomes}

The three core dispositions illustrated in Fig. \ref{fig:dispositions} within the Prompt-to-Primal (P2P) Teaching framework are measurable and can serve as reliable indicators of the overall effectiveness of the teaching–learning process. For instance, the proportion of students participating in the prompting phase, together with the number of additional questions (beyond the immediate questions from the topics provided by the instructor) reflect the cohort’s level of intellectual curiosity; in the present study, 41\% of students engaged in Phase 1. 

Concerning the subsequent dimension represented in the three-dimensional core disposition graph, a comparative analysis between classes conducted with and without the P2P methodology demonstrated a 38\% increase in the number of in-class questions when P2P was implemented, indicating a significant enhancement in the Engagement dimension. It is noteworthy that, given students’ prior exposure to general conceptual information through the LLM-based dialogue, not only did the number of questions increase, but their quality also improved. The questions were more explicitly oriented toward first-principles reasoning. This qualitative shift in inquiry serves as a valuable indicator of deeper cognitive engagement and conceptual understanding.  

The student perception of the AI-assisted learning approach was evaluated using a questionnaire administered to the participating students. The student assessment results reveal a strong central tendency toward positive educational outcomes and critical skepticism regarding AI tools. Regarding Conceptual Alignment, the distribution was highly skewed, with the modal response being \enquote{Mostly aligned} at a commanding 75\%. This indicates the AI-assisted understanding largely cohered with the first-principles model taught in class, with only a quarter of the cohort reporting partial or mostly different alignment and none reporting the extreme outcomes of perfect or complete misalignment. The process of reconciliation clearly boosted learning, as evidenced by the results for Depth of Understanding: the modal response was \enquote{Moderately improved} at 50\%, with a combined 62.5\% of students reporting moderate or significant improvement. However, a notable 25\% reported \enquote{No change,} suggesting the reconciliation provided limited marginal utility for a quartile of the sample. In terms of Critical Validation Skill, the results showed a universally positive shift in self-reported confidence; the modal response was \enquote{Somewhat confident} at 62.5\%, with the remaining 37.5\% selecting \enquote{Neutral,} and 0\% selecting any uncertain or not confident categories. This suggests the exercise successfully cultivated a fundamental belief in the ability to evaluate AI outputs. This cautious view is reinforced by the data on AI as a Learning Partner, where the modal response was \enquote{Mostly requires verification} at 50\%, and a total of 62.5\% adopted a verification-centric view, contrasting sharply with the 12.5\% who saw AI as \enquote{Mostly exploratory.} Finally, the Subsequent Phase revealed a strong willingness for manual reinforcement, with the modal response being \enquote{Yes} at 62.5\%, and no students outright refusing the follow-up manual review. 

Following this internal validation, the quantitative results underscore the methodology's positive impact on academic performance. The average score of the midterm exam increased by 11\% with the adoption of the P2P teaching and critical AI-validation approach, suggesting that the students' increased conceptual alignment and critical engagement translated directly into measurable learning gains.

\section{Conclusion}

The Prompt-to-Primal (P2P) framework offers a robust, modern solution for integrating generative AI into engineering education. This model functions as a unique hybrid, pairing student-driven AI discovery with disciplined first-principles reasoning. The core innovation of P2P lies in its deliberate treatment of AI as a fallible pedagogical tool , actively using AI-generated output, including its occasional errors—as the focal point for critical examination and classroom discussion. This approach directly combats the passive learning that leads to the \enquote{illusion of understanding.} By consistently embedding AI interaction within a structured, instructor-mediated process, P2P not only cultivates interpretive rigor and responsible AI literacy , but also transforms AI from a cognitive risk into a structured learning amplifier. The P2P framework is therefore a scalable and adaptive model for building deep, foundational knowledge in the age of artificial intelligence.  

\bibliographystyle{IEEEtran}

\begin{thebibliography}{99}

\bibitem{lee2025} D.~Lee and E.~Palmer, \textit{Prompt engineering in higher education: a systematic review to help inform curricula}. International Journal of Educational Technology in Higher Education, 2025.

\bibitem{qian2025} Y.~Qian, \textit{Prompt Engineering in Education: A Systematic Review of Approaches and Educational Applications}. Journal of Educational Computing Research, 2025.

\bibitem{carrasco2025} J.~L.~Carrasco-Sáez, C.~Contreras-Saavedra, S.~San-Martín-Quiroga, C.~E.~Contreras-Saavedra, and R.~Viveros-Muñoz, \textit{Analyzing Higher Education Students' Prompting Techniques and Their Impact on ChatGPT's Performance: An Exploratory Study in Spanish}. Applied Sciences, 2025.

\bibitem{ward2025} B.~Ward, D.~Bhati, F.~Neha, and A.~Guercio, “Analyzing the Impact of AI Tools on Student Study Habits and Academic Performance,” in *Proc. IEEE 15th Annual Computing and Communication Workshop and Conference (CCWC)*, 2025, pp. 434-440.

\bibitem{ciolacu2024} M.~I.~Ciolacu, C.~Marghescu, B.~Mihailescu, M.~Sorecau, E.~Sorecau, and P.~Bechet, ``Education 5.0: Transforming Engineering Education in the Age of Generative AI,'' in \textit{2024 IEEE 30th International Symposium for Design and Technology in Electronic Packaging (SIITME)}, Sibiu, Romania, 2024, pp.~303--308.

\bibitem{guedes2025} P.~Guedes, E.~Abranches~Silva~Lopes, P.~F.~Ribeiro, and A.~C.~Zambroni~de~Souza, ``The Impact of Artificial Intelligence on Learning and Teaching of Engineering,'' \textit{IEEE Transactions on Education}, vol.~68, no.~5, pp.~417--425, 2025.

\bibitem{hao2025} Y.~Hao, Y.~Liu, B.~Liu, G.~Amarantidis, and R.~Ghannam, ``Integrating AI in Engineering Education: A Comprehensive Review and Student-Informed Module Design for U.K. Students,'' \textit{IEEE Transactions on Education}, vol.~68, no.~2, pp.~173--185, 2025.

\bibitem{novak1999} G.~Novak, E.~Patterson, A.~Gavrin, and W.~Christian, \textit{Just-in-Time Teaching: Blending Active Learning with Web Technology}. Upper Saddle River, NJ, USA: Prentice Hall, 1999.

\bibitem{marzano2001} R.~J.~Marzano, D.~J.~Pickering, and J.~E.~Pollock, \textit{Classroom Instruction That Works: Research-Based Strategies for Increasing Student Achievement}. Alexandria, VA, USA: Association for Supervision and Curriculum Development, 2001.

\bibitem{bishop2013} C.~Bishop and M.~Verleger, ``The flipped classroom: A survey of the research,'' in \textit{Proc. ASEE Annu. Conf.}, 2013.

\bibitem{karagianni2024} G.~K.~Karagianni, \textit{Metacognitive Evolution: Bridging Aristotelian Wisdom and Autonomous Learning in the Digital Age}. International Journal of Education and Research, 2024.

\bibitem{zhai2024} C.~Zhai, S.~Wibowo, and L.~D.~Li, ``The effects of over-reliance on AI dialogue systems on students' cognitive abilities: a systematic review,'' \textit{Smart Learning Environments}, vol.~11, art.~28, 2024, doi: 10.1186/s40561-024-00316-7.

\bibitem{chi2009} M.~T.~H.~Chi, ``Active-constructive-interactive: A conceptual framework for differentiating learning activities,'' \textit{Topics in Cognitive Science}, vol.~1, pp.~73--105, 2009.

\bibitem{messeri2024} L.~Messeri and M.~J.~Crockett, ``Artificial intelligence and illusions of understanding in scientific research,'' \textit{Nature}, vol.~627, no.~8002, pp.~49--58, 2024.

\bibitem{bruner1960} J.~Bruner, \textit{The Process of Education}. Cambridge, MA, USA: Harvard Univ. Press, 1960.

\bibitem{flavell1979} J.~Flavell, ``Metacognition and cognitive monitoring: A new area of cognitive–developmental inquiry,'' \textit{Amer. Psychologist}, vol.~34, no.~10, pp.~906--911, 1979.%5

\bibitem{brooks1999} J.~G.~Brooks and M.~G.~Brooks, \textit{In Search of Understanding: The Case for Constructivist Classrooms}, 2nd ed. Alexandria, VA, USA: Association for Supervision and Curriculum Development (ASCD), 1999.%6

\bibitem{prince2006} M.~J.~Prince and R.~M.~Felder, ``Inductive Teaching and Learning Methods: Definitions, Comparisons, and Research Bases,'' \textit{Journal of Engineering Education}, vol.~95, no.~2, pp.~123--138, 2006.

\bibitem{wilson2000} B.~G.~Wilson and K.~M.~Myers, \textit{Situated cognition in theoretical and practical context}. Theoretical Foundations of Learning Environments, 2000.

\bibitem{saville2011} K.~Saville, \textit{Strategies for using repetition as a powerful teaching tool}. Music Educators Journal, 2011.

\bibitem{agarwal2021} P.~K.~Agarwal, L.~D.~Nunes, and J.~R.~Blunt, ``Retrieval Practice Consistently Benefits Student Learning: a Systematic Review of Applied Research in Schools and Classrooms,'' \textit{Educ Psychol Rev}, vol.~33, pp.~1409--1453, 2021.

\bibitem{vanderweel2024} F.~R.~Van der Weel, and A.~L.~H.~Van der Meer, ``Handwriting but not typewriting leads to widespread brain connectivity: a high-density EEG study with implications for the classroom,'' \textit{Frontiers in Psychology}, vol.~15, Art.~1219945, 2024.

\bibitem{mollick2024} A.~Mollick and E.~Mollick, ``Using AI to implement effective teaching strategies in classrooms,'' \textit{Comput. Educ.: Artif. Intell.}, vol.~5, 2024.

\bibitem{jonassen1999} D.~Jonassen, ``Designing constructivist learning environments,'' in \textit{Instructional Design Theories and Models: A New Paradigm of Instructional Theory}. Mahwah, NJ, USA: Lawrence Erlbaum, 1999.


\end{thebibliography}

\end{document}